# Law of Genome Evolution Direction :

# Coding Information Quantity Grows

**Liaofu Luo**

**Laboratory of Theoretical Biophysics, Faculty of Science and Technology, Inner Mongolia University, Hohhot 010021, China;**

*Email address: lolfcm@mail.imu.edu.cn

## Abstract

The problem of the directionality of genome evolution is studied. Based on the analysis of C-value paradox and the evolution of genome size we propose that the function-coding information quantity of a genome always grows in the course of evolution through sequence duplication, expansion of code, and gene transfer from outside. The function-coding information quantity of a genome consists of two parts, p-coding information quantity which encodes functional protein and n-coding information quantity which encodes other functional elements except amino acid sequence. The evidences on the evolutionary law about the function-coding information quantity are listed. The needs of function is the motive force for the expansion of coding information quantity and the information quantity expansion is the way to make functional innovation and extension for a species. So, the increase of coding information quantity of a genome is a measure of the acquired new function and it determines the directionality of genome evolution.

The extension of the diversity of species and their evolution towards higher function more adaptive to environment shows that the life evolution obeys a law with definite direction. Darwin expressed the law as “survival of the fittest”. It means that the ‘designed’ properties of living things better adapted to survive will leave more offspring and automatically increase in frequency from one generation to the next, while the poorly adapted species will decrease in frequency. Accompanying the development of molecular biology and with ever increasing understanding on genomes we are able to express the law on life evolution more quantitatively and precisely. Here the key point is the introduction of information. Life consists of matter and energy, but it is not just matter and energy. The life of an individual comes from the DNA of its parents. DNA, which weighs only $10^{-12}$ gram for human, is insignificant in terms of matter because, like many other things on earth, it is composed of nitrogen, oxygen, sulfur, etc. In addition, DNA, as a source of energy, is also unimportant, since it is just composed of the similar level of chemical energy as other macromolecules that can be produced by experiment. However, different from matter and energy, information constitutes the third fundamental category in natural sciences. Schrodinger (1944) was the first who recognized the importance of information and indicated that the characteristic feature of life which differentiates from an inanimate piece of matter is the large amount of information contained in its chromosomes. He said, “We believe a gene – or perhaps the whole chromosome fibre – to be an aperiodic solid. With the molecular picture of gene it is no longer inconceivable that the miniature code should precisely correspond with a highly complicated and specified plan of development and should somehow contain the means to put it into operation." Sixty years have passed. Now, as we try to formulate the law on the directionality of life evolution we should put our discussion based again on the concept of information. Thanks to the discovery of cell totipotent, despite the complexity of multicellular organnism the full genetic information of an organism is stored in chromosomes of just one cell. We can express the basic law on species evolution through the genomic information contained in chromosomes of one cell. Many studies were carried out about the evolution of genome size (Gregory et al, 2007). The size of genomes for sibling species can change more than several tenfold, for example, 340-fold for flatworms, 70-fold for nematodes, 170-fold for arthropods (insecta), 350-fold for fish, 130-fold for amphibians,196-fold for algae (chlorophyta), 500-fold for pterridophytes and 1000-fold for angiosperms, etc. Moreover, the evolutionary complexity of a species is irrespective of its genome size. For example, the C-values of lungfishes are higher than human about 8 to 20 times. Some salamanders can also have large genomes with C-values 15 times or more than human. Next, about the genome size evolution it was speculated that plants might have a "one-way ticket to genomic obesity" through amplification of retrotransposons and polyploidy. The similar assumptions were also made that the animal genome sizes might change in the direction of increase. However, there has been considerable evidence that both increases and decreases may occur in plant and animal lineages. In the meantime, no fossil evidence on the genome size variability in the single direction was reported (Gregory, 2005). These have made the issue on the directionality of genome size more complex and interesting. However, from our point of view the genome size of a species is not a proper measure of evolutionary directionality since it is irrespective of genetic information necessary for encoding the biological function. Instead, when our investigation is based on the function-coding information we will be able to obtain an unambiguous picture on the evolutionary law of genomes

# 1. The law of genome function-coding information quantity growing

Definition. For an n- long sequence (called sequence A) written by symbols $A_1$, $A_2$,…,.and $A_n$ where $A_i$ taking $s_i$ possible values ($i$=1,…,n) , if the sequence A encodes certain function then we define the function-coding information quantity of the sequence $I_C = \log_2 \prod_i s_i$ .

For prokaryote genome the symbol $A_i$ takes four values A,G,C or T. For eukaryotes, the chromatin remoulding and histone modifications is another type of variables which has the potential to influence fundamental biological processes and may be epigenetically inherited. More than thirty histone modifications have been found for human and other vertebrates (Kouzarides, 2007). They form the source of heredity information in addition to four kinds of bases A,G,C and T. Moreover, the DNA methylation (mainly the cytosine methylation for higher vertebrates) can inhibit gene expression and be epigenetically inherited and therefore the methylated bases give additional symbols applicable in genetic language (Lewin, 2008).

Following the definition，a function-coding DNA sequence segment (or its extension which includes the chromatin remoulding variable and methylated cytosine et al ) consists of two basic types：p-coding sequence which encodes functional protein and n-coding sequence which encodes all other functional sequence segments.

**Law of Function-Coding Information Quantity Growing (CIQG)** Due to the interaction among DNA,RNA and protein under the possible influence of chromatin remoulding, chemical modification and other factors the function-coding information quantity $I_C$ of a genome sequence always grows in the course of evolution $(\frac{dI_C}{dt} \geq 0)$ through sequence duplication, expansion of code, and horizontal gene transfer.

The evolutionary law was firstly proposed in references (Luo, 2005, 2006) as an assumption. Before its demonstration we shall give some comments on the expression of the law:

1, Entropy increase is a universal law of nature. Due to the randomness of stochastic movement the entropy of any isolated physical system always increases. However, what stated here is not identical with the physical law of entropy growing. The law of CIQG refers to the evolution of function-coding information quantity of genome. Here the mechanism responsible for the law is functional selection. As a biological law of evolution, there may exist exception; for example, some genome evolves in strange and peculiar environment. However, the exception should be very rare. On the other hand, the time scale *dt* in $\frac{dI_C}{dt} \geq 0$ is determined by the minimal time interval required for natural selection acting on heredity process. For a given specie as the clock with accuracy of several generations is used the law $\frac{dI_C}{dt} \geq 0$ will manifest itself.

2, Evolutionary law is closely related to the environment. Recently, the evidences for environmental change as a determinant of mass extinction and its biological selectivity in the fossil record were indicated (Peters, 2008). For a genome in stable environment $\boldsymbol{I_C}$ grows with time. For a genome in varying environment $\boldsymbol{I_C}$ grows adapting to the change of environment. The speed of coding information quantity growing can be viewed as a scale of evolutionary rate of species. However, for a genome in suddenly-changing environment, if the speed of growing $\boldsymbol{I_C}$ cannot adapt to the sudden change of environment (for example, the food deficiency), then the species would be close to extinction; or new species would otherwise emerge, the genes of which can adapt to the functional needs under new environment. From fossil records we know that evolution has a range of rates, from sudden to smooth and both punctuated equilibrium and phyletic gradualism have occurred (Ridley, 2004). The punctuated equilibrium may be somewhat commoner than phyletic gradualism and in new species formation the evolution always shows a sudden rate. Whatever the change is in sudden or smooth phase the law of CIQG holds universally.

3, The loss of function in parasitism of some bacteria resulting in the decrease of coding information quantity of these genomes is a phenomenon of retrogression, which should not be included in the scope of the law. The law holds only for free-living genomes. The environment inside the host cell contains many of the nutrients and defense systems that a bacterial cell needed, since genes that are needed in a free-living bacterium to provide the resources are not needed in an intracellular bacterium. In this case the gene loss may be advantageous since a cell with less DNA can reproduce faster.

## 2. Supporting facts in demonstration of CIQG law

1) Is the law of function-coding information quantity growing for a genome consistent with experimental data? Biologists know that the evolutionary complexity of a species is irrespective to its genome size. But, as plotting the range of genome size in different evolutionary phyla, we found that there is an increase in the minimum genome size in each group as the complexity increase. (Lewin, 2004, 2008). This is due to the quantity of proteins in different species of a group basically same and the size of minimum genome reflecting the p-coding information quantity, while the p-coding information quantity correlated well to the total function-coding information quantity from prokaryotes to lower eukaryotes.

2) The prokaryote genome size changes from 1Mb to 10 Mb for free-living bacteria. The variation is relatively smaller than for eukaryotes. Because there is so little noncoding DNA in prokaryotes the mechanism for generating the variation must largely involve changes in gene number. Both gene number increase and reduction over the course of prokaryotes evolution are observed (Gregory & DeSalle, 2005). However, the gene loss has happened primarily among parasites and symbionts. In the meantime, the shift of a genome from free-living to host-dependent lifestyle renders many genes obsolete. They are no longer maintained by selection, but deactivated by mutation and then subsequently lost. The process may be long and it leads to pseudogenes remaining intact in some genomes (Mira et al, 2001). Since DNA deletion is mainly related to parasites and pseudo-genes the free-living bacterium always shows evolutionary directionality towards the increase of function-coding information quantity. The consistency of CIQG law with data of prokaryote genome evolution can be proved by the following observation. Suppose the phylogenetic relations among prokaryotes have been known and a tree of life for free-living bacteria has been reconstructed. The terminal nodes of the tree represent the present species and the internal nodes represent the past-time genomes. The coding information quantity of a present bacterium can be deduced directly by calculating the functional gene number in the genome.which equals the total number of .genes minus the number of pseudogenes. While for ancestor species it should be obtained based on some assumptions. Since only the free-living bacteria are concerned, as a working hypothesis we suppose that the genome size of an ancestor species takes the minimal value of sizes of its first descendents in the clade. Thus we are able to obtain a self-consistent solution for the coding information quantity of all free-living bacteria. We find the solution satisfying the law of coding information quantity increasing in evolution. Note that due to the functional diversity some genomes on the tree, for example, *Bradyrhizobium*, *Streptomyces* et al possess size much higher than their sibling branches. But, regardless of the large differences in genome size, we have shown the CIQG law consistent with prokaryotic genome data.

3) For eukaryotes the DNA loss was also frequently observed in genome evolution. However, the lost DNA is generally related to non-function region. There has been a proposition that at the level of small (<400bp) insertions and deletions (indels) the deletions in a genome are more frequently occurred than the insertions. Petrov et al (2000) studied the indel spectrum in *Laupala* crickets and in *Drosophila*. The former has a genome size 11 times larger than latter. They found the DNA loss in *Laupala* is more than 40 times slower than in *Drosophila* and indicated the possible inverse correlation between genome size and DNA loss rate. However, the DNA losses discussed by these authors are mainly related to pseudogenes which lack the capacity to encode functional proteins. So, they successfully explained that the high rate of DNA loss in small genomes results in a lower steady-state number of pseudogenes. The result does not mean at all that the loss of functional DNA has happened in some eukaryotic genomes. Generally speaking, gene duplication and loss is a powerful source of functional innovation. Wapinski et al (2007) studied the evolutionary principles of gene duplication in fungal genomes through determination of orthology and paralogy relations across 17 species and demonstrated that gene duplication and loss is highly constrained by the functional properties and interacting partners of genes. They indicated that the duplicated genes typically diverge with respect to regulatory control and therefore the gene duplication may drive the modulalization of functional networks through specialization. This means the trend of the function-coding information quantity increasing in evolution.

4) An important approach of new species formation is the whole genome duplication (WGD) which leads to several new species by different ways of gene-loss after the duplication. For example, Scannell and Byrne et al (2006) studied reciprocal gene loss in polyploidy yeasts. A whole-genome duplication occurred in a shared ancestor of yeast species *S cerevisiae*, *S castellii* and *C. glabrata*. These authors traced the subsequent losses of duplicated genes and showed that the pattern of loss differs among three species at 20% of all loci. Although three species lose genes respectively in their divergence, the total amounts of genes increase due to genome duplication, the increment 14％ for *S cerevisiae*, 16％ for *S castellii* and 11％ for *C glabrata*. The pattern of reciprocal gene loss demonstrated the mechanism of reproductive isolation and the further analysis indicated the rapid divergence of three species shortly after the whole-genome duplication. Aury et al (2006) studied the whole-genome duplications in ciliate. These authors reported that most of the nearly 40000 genes of the unicellular eukaryote *P. tetraurelia* arose through at least three successive whole-genome duplications and the most recent duplication gave rise to the *P. aurelia* complex of 15 sibling species. They observed that the gene loss occurs over a long timescale and the temporary maintenance of many duplicated genes is simply due to dosage constraints. Then, they estimated the gene number of the species about 20000 in the initial phase, from 20000 to 40000 in first WGD, from 21000 to 42000 in second WGD, from 26000 to 52000 in third WGD and 39642 after the third genome duplication and speciation. The series of gene number 20000, 21000, 26000 and 39462 approximately reflect the increase trend of functional gene. So, in spite of a large amount of gene loss in WGD which was observed in above two examples the total number of functional genes in a genome still increases through many gene duplications. This is consistent with the law of coding information quantity growing in evolution. In fact, the gene duplication is associated with the acquisition of some new gene function. The new function associated with gene duplication can arise in different ways, for example, neo-functionalization and escape from adaptive conflict etc. (Des Marais, Ransher, 2008). Here the essential point behind apparently different approaches is the increase of coding information quantity which can serve as a measure of the acquired new function for a genome.

5) It was estimated that 70% of all angiosperms had experienced one or more episodes of polyploidy in their ancestry. Intuitively, the polyploids should have larger C-values than diploids with its C-value increasing in direct proportion to ploidal level. However, the loss of DNA following polyploid formation may be a widespread phenomenon (Leitch & Benett, 2004). There are three potential outcomes for the plant genes duplicated by polyploidy: first, both copies remain functional; second, one copy becomes silences or lost while the other retains the original function; or third, the two copies may diverge in function. In the second possibility the gene loss can occur rapidly. For example, recent studies have demonstrated the polyploidy-induced rapid and reproducible elimination of DNA sequences in the wheat group. But the eliminated sequences were not homologous to known genes and likely represented noncoding regions (Ozkan et al, 2001).

6) In the Cambrian explosion of animals (metazoan expansion) and in the fast evolution from reptiles to birds and to mammals one can find more evidences on CIQG law from these adaptive radiation events. An adaptive radiation means that a small number of ancestral species in one taxon diversifies into a large number of descendant species, occupying a broader range of ecological niches. Darwin was interested in why evolution usually shows a diverging, tree-like pattern. He explained the pattern by competition. More similar forms will compete more strong, which tends to push species apart during evolution. The proliferation of animals with hard skeletons in the Cambrian explosion may be explained by predators evolving escalated skills around this time. The new function needs the increase of information quantity to encode it. The animals with hard parts may have higher coding information quantity than their soft-bodied predecessors (Ridley, 2004). The origin of mammals can be traced back before 85-100 million years ago, through a series of small changes in mammal-like reptiles (Bininda-Edmonds et al, 2007). The reported reptilian C-values vary from 1.1 pg to 5.4 pg while the mammalian C-values range from 1.7 pg to 8.4 pg (Gregory, 2005). Considering the C-values measured only for a small fraction of modern species and the extinction of many evolutionary predecessors, the C-value data given above is far from complete. If the coding information quantity can be estimated by the lower bound of the observed C-values in an appropriate group of species then the evolution from reptiles to mammals seems not inconsistent with CIQG law. The origin of bird flight is another important event in vertebrate evolution. The reported avian C-values range from 1.0 pg to 2.2 pg for 2% studied species (Gregory, 2005). The minimal C-values observed in birds are close to those of reptiles. Adapted to the flight condition the birds should have a relatively small genome. Recently, some fossil evidences showed that the birds evolve from non-avian dinosaurs, the saurischian dinosaur lineage (Xu et al, 2004). The latter has less repetitive DNA than other typical ancestral Dinosauria. These genomic characteristics of fewer repetitive elements and less non-coding DNA should be added to the list of attributes previously considered avian but now thought to have arisen in non-avian dinosaurs, such as feathers, pulmonary innovations and parental care and nesting (Organ et al, 2007). They exhibited a structure of preadaptation which happened to evolve new function of flight. Although we don’t know the details of genome variation yet about the acquisition of avian flight function the fast evolution from reptiles to birds should not be in conflict with CIQG law, too.

7）Both increase and decrease of genome size occur in plant and animal lineages. The DNA amount of a genome reflects the dynamic balance between the opposing forces of expansion and contraction. However, the steady state changes adiabatically (more slowly than the force acting on genome) towards the higher functional state through selection. We shall investigate the contraction forces in more detail. The mechanism of sequence reduction induced by contraction forces include mainly the sequence recombination by unequal cross-over during meiosis and unequal sister chromatid exchange during mitosis. The process of unequal intrastrand homologous recombination occurs between the long terminal repeats of LTR-retrotransposons, and can lead to deletion of the internal DNA segment and one LTR, leaving behind only one ‘solo LTR’. In this case, there is still a net gain of DNA with each insertion of transposable element (Bennett & Leitch, 2005). On the other hand, the illegitimate recombination, irrespective of homologous sequences, can act on a larger fraction of the genome. It may involve the deletion of all intervening sequences between the two LTR elements. It was indicated that the illegitimate recombination is the main driving force behind genome size decrease in *A. thaliana* (Devos et al, 2002). The repair of double stranded breaks in DNA is another important mechanism responsible for sequence deletion observed in some plants. The comparison between *A. thaliana* and tobacco genomes shows there exist marked differences in their repair pathways after double stranded breaks (Filkowski et al, 2004). In all of these mechanisms of sequence reduction, we speculate that the lost DNA segments may be redundant, not functionally important or, though being functional elements, they have been deleted and replaced by a more efficient functional network through repair pathway as observed in the case of double stranded breaks. So, the sequence reductions are not likely to contradict the CIQG law.

8) The p-coding information quantity grows slowly in higher eukaryotes, not proportional to the increase of genome complexity. For example, the sequence length of protein-coding DNA in human genome is nearly same as in mouse and even in all vertebrates. On the other hand, we know that the gene density in a genome decreases explicitly with the growing evolutionary complexity of species. The gene density is 1000 genes per Mb for prokaryotes to 500 genes per Mb for yeast, and 20 genes per Mb for mammals. This indicates the complexity of regulation mechanism increasing with evolution. So we have to consider n-coding information quantity in addition to p-coding information quantity. In fact, the complexity of a genome originates from the gene function, which is determined mainly not by the number of genes contained in it, but by the interaction among genes and the gene regulation. The comparison between human proteome and other eukaryotes shows that most protein domains appears to be common to the animal kingdom. However, in human there are many new protein architectures – new combinations of domains. The greatest increase occurs in transmembrane and extracellular proteins, which may be related to the addition of functions required for the interaction between the cells of a multicellular organism (Lewin, 2004, 2008). Recently, the ENCODE project analyzed the functional elements in 1% of the human genome. In the ‘constrained sequences’ (genomic sequences orthologous to the ENCODE regions from 14 mammalian species) that serve some functional roles and correspond to 4.9% of the nucleotides in the studied regions they found the protein-coding sequences only amount to 32% (ENCODE, 2007). This indicates that in the functional region of human genome a large portion of sequences are not responsible for protein-coding but possibly for gene regulating. Consider the gene expression regulation at transcriptional level. The rate of transcription is modified by the binding to enhancer sequences or other regulatory elements (RE) by transcription factors (TF). REs are typically 5 to 10 bp sequences, often assembled into regulatory modules (enhancers, repressors, etc). A module typically contains binding sites for 4 to 8 TFs. Here the complexity consists in: related TFs may recognize similar binding motifs (so there can be cross reaction), and the RE sequences for a given TF can vary (variation in recognition sequences can lead to binding by different TFs) and the location of REs also varies among genes and among species (on either side of the gene near the gene or tens of kilobases away). So, the number and nature of regulatory elements are easily changed in a reasonably short periods of evolutionary time. Therefore, even if the protein-coding part of a gene is not altered, its usage may be. The information contained in regulatory mechanism is n-coding information quantity. Taking the variability of RE into account, if there are $4^{10}$ REs of 10 bp and each RE occurs once in genome then it should amount to $1.68\times10^8$bp, about 1/20 of human genome. The above estimate only accounts for the transcriptional regulation. The gene regulation at other levels needs more coding information quantity. So, one may conclude with assurance that for higher organisms the n-coding information quantity is much larger than p-coding information quantity in a genome. The sum of p- and n-coding information quantity will be able to show a stronger correlation with genome evolution.

Many discussions on junk DNA were carried out in recent years. “Not junk after all,” people said.(Makalowski, 2003; Wickelgren, 2003). From ultraconserved non-genic DNA sequences in mammalian genomes (Dermitzakis et al, 2003; Bejerano et al, 2004), to non-coding RNAs (such as micro RNAs and small interference RNAs, cf Storz, 2002; He et al, 2004) and to mobile elements (Kazazian, 2004) — all these elements may code for certain functions. They should have contributed to n－coding information quantity. Recent studies on human chromosome 18 provide a new example of functional role of non-protein-coding elements (Nusbaum et al, 2005). Despite the low density of protein-coding genes on chromosome 18, it has been found that the proportion of non-protein-coding sequences evolutionarily conserved among mammals is close to the genome-wide average. These sequences might serve a structural role, with a constant density of such elements required to maintain chromosome structure independent of gene density.

9) The law of coding information quantity growing has found more direct evidence in the evolution of human genome (Liu et al, 2003). Through comparison of human genome with chimpanzee, baboon and lemuer, Liu and Eichler et al found a 15% to 20% expansion of human genome size over the last 50 million years of primate evolution. More plainly, these three species —chimpanzee, baboon and lemuer— are estimated to have diverged from human at three different time point, approximately 5.5, 25 and 55 million years ago; and compared to chimpanzee and lemuer the human genome is estimated to have expanded 30 Mb, and 550 Mb respectively. Under the assumption that the genome size of lemuer, baboon and chimpanzee increases slowly respectively after their divergence and the expansion of human genome is related to certain events coding for new functions, the following picture can be deduced: the coding information quantity of human genome (including p-coding and n-coding sequence) has grown 30Mb in last 5.5 Myr, and 550Mb in last 55 Myr. In fact, orthologous comparisons have shown that 90% of the human expansion is due to new retroposon insertions and it is reasonable to assume that the insertion and fixation of retroposons L1 and Alu in human genome are related to the emergence of new functions. For example, BC200, a brain-specific RNA that is part of a ribonucleoprotein complex preferentially located in the dendrites of all anthropoid primates, appears to have been derived about 35-55 million years ago from an Alu transposable elements (Smit, 1999). Many of the genetic differences between humans and other primates are resulted from transposable element activity —through the action of regulatory regions descended from transposable elements (TE) and/or via TE-mediated exon splicing and deletion mutations. The inactivation of the CMP-N-acetylneuraminic acid hydroxylase enzyme gene in humans is accomplished by an exonic deletion caused by the insertion of a human-specific Alu elements. This deletion mutation occurred prior to brain expansion during human evolution (Chou et al, 2002). Sometimes, the deletion of an old gene is favorable to gain a new gene or to form some new biochemical pathways for advanced function. The total coding information quantity of the genome is increased in the process. It was estimated that total 700-1800 duplications have occurred in the human genome since the split with chimpanzees. These duplications and accompanying some deletions of genes may generate the variation in the gene expression between chimps and humans, especially in the brain (Enard et al 2002). The loss of body hair is another example of the role of gene deletion in human evolution. The ape hair keratin is functional and encoded in clustered gene families, meaning that they were produced by duplication, while in humans a type I hair keratin pseudogene inactivated by a single point mutation was observed (Winter et al, 2001).

## 3. Approaches to information quantity growing in a genome

There are four mechanisms responsible for the expansion of coding information quantity in a genome, namely, the sequence duplication that increases the genome size, the functionalization of transposable elements and other ‘junk’ DNA, the formation and employment of diverse modes for coding, and the gene horizontal transfer from other species.

The sequence duplication includes global duplications (in which entire genome or a chromosome is duplicated) and regional multiplication. The distribution of bacterial genome size is discontinuous, showing major peaks at around 0.8Mb, 1.6Mb, and 4.0Mb. This distribution has led to the hypothesis that the larger genomes evolved from smaller ones by genome duplication. The polymodal distribution of genome sizes in many groups of eukaryotes (for example, in monocotyledon plants) also suggests that polyploidy is a major mechanism in eukaryotic evolution. There is strong evidence that two rounds of whole-genome duplication occurred in the vertebrate evolution. The regional multiplications are more frequently observed in many genomes. About 38%-45％ of genes in *E coli* genome are identified as paralogous families arisen from gene duplication; about 47% of genes in *B subtilis* genome constitute paralogous gene families, etc.. In human genome about 53% of DNA is occupied by repetitive sequences. The localized regional increase of genomic DNA through tandem duplications may be created by replication slippage, slipped-strand mispairing, hairpin formation, etc.. Replication slippage provides a powerful mechanism for the rapid proliferation of tandemly repeated sequences within a genome. In the case of long repeated units, the major mechanism for expansion is unequal crossing–over or DNA amplification due to multiple replication of the same replicon (Patthy, 1999).

The repetitive DNA sequences constitute a large portion of the genomes of eukaryotes. They arise from sequence duplication. The ‘selfish DNA’ hypothesis proposes that they are maintained by their ability to replicate within the genome. They are ‘junk’ produced in neutral mutation. For example, the inverse correlation between development rate and amount of highly repeated DNA in flowering plants and salamanders is compatible with non-selective mechanisms for maintaining repeated arrays since more slowly developing species could accumulate larger quantities of repetitive sequences generated by array size expansion. The large differences in amounts of satellite sequences between closely related species and the absence of measurable fitness effects of large deletions or duplications of heterochromatin in *Drosophila* also support non-selective mechanisms (Charlesworth et al, 1994). So, there is a considerable proportion of neutral elements (repetitive DNA sequences) that do not confer a selective advantage or disadvantage to the organism. However, the neutral non-selective mechanism for the maintenance of repetitive sequences does not mean they are genomic garbage of completely no use. In fact, genomes are dynamic entities: New functional elements appear and old ones become extinct. The neutral pool of sequence elements may turn over during evolutionary time, emerging via certain mutations and disappearing by others. Those sequences having not been eliminated by evolutionary process might be related to their acquisition of some new biological role. The evolutionary picture is: some repetitive sequences acquire new functions and form new genes or regulatory elements, making the increase of coding information quantity; while some ones lose the activity and change to pseudogene, without contributing to the coding information quantity in the further evolution. Above points were indicated by several authors in the last decade, including by a Chinese monograph published in 2000 (Luo, 2000). Now, more and more biologists recognized the importance of functionalization of ‘junk’ DNA and regarded the neutral pool of sequence elements as a genomic treasure (ENCODE, 2007).

The most important examples of the functionalization of neutral pool elements are retrotransposons. The mobile elements (transposable elements) comprise a group of distinct DNA sequences that have the ability to integrate into the genome at a new site. These elements include DNA transposons, LTR (long terminal repeats) retrotransposons, non-LTR autonomous retrotransposons and non-LTR nonautonomous retrotransposons. DNA transposons occupy about 3% of human genome. These elements are generally excised from one genomic site and integrated into another by a ‘cut and paste’ mechanism. Retrotransposons are transcribed and reintegrated into the genome, thereby duplicating the elements. LTR retrotransposons is similar to retroviruses in structure. Non-LTR retrotransposons are classified into autonomous (with ability to encode endonuclease and reverse transcriptase) and nonautonomous (not encoding endonuclease and reverse transcriptase). The former is typified by LINE-1 (long interspersed nucleotide elements-1) of mammals and the latter typified by SINE (short interspersed nucleotide elements) or Alu sequences in human. Both they are highly-repetitive sequences. The amount of retrotransposons in a genome increases with evolution, about 3%-5% for lower eukaryotes but close to 50% for mammals. LINE-1 sequences have accumulated to 17% of the human genome while Alus have expanded to 1.1 milllion copies or 11% of it. These elements have driven genome evolution in diverse ways: they are drivers of genome evolution (Kazazian, 2004). Mammalian L1 elements affect the genome in many ways, for example, repair of double-strand breaks, expression of genes 5’ to full-length L1s via an antisense promoter in L1, use of L1 in coding regions of genes, etc. A large burst of Alu insertion has happened 40 million years ago in hominid lineage but after that the activity declined markedly. Alu elements modulate gene expression at the post-transcriptional level in at least three independent manners: alternative splicing, RNA editing and translation regulation. So, the fast evolution of vertebrates has put the transposable elements to use. Based on three-element-interaction among DNA, RNA and protein the retrotransposons, carrying a large amount of information, should be regarded as a large reservoir of regulatory functions that have been actively participating in mammalian evolution. Especially, the L1 and Alu sequences may have played distinct roles in the genome evolution of hominid lineage.

New code formation is another important origin of information quantity growing in a genome. The employment of diverse modes for coding leads to the expansion of genome coding information quantity. The formation of code in a complex system is caused by the stochastic interaction among its sub-systems. It emerges from the structural matching of subunits and the physico-chemical interaction among them. Once the new mode for coding being formed and selected for functional use the mode will survive in evolution. The emergence of $21^{st}$ amino acid selenocysteine and the $22^{nd}$ amino acid pyrrolysine are two examples. Both they are produced from the formation of new coding rules by reinterpreting nonsense codons (Stadtman, 1996). We emphasize that the code used in genome is far from only one single mode, the amino acid code.

For instance, microRNA (miRNA) and small interfering RNA (siRNA) provide examples of diverse codes other than amino acid code. Both they are 21-25 nucleotide non-coding RNAs which regulate gene expression in a sequence-specific manner. These molecules are recognized and processed by a common RNase-III processing enzyme – Dicer. They are assembled into the RNA-induced silencing complex (RISC). The effector complex RISC subsequently acts on its target by translational repression or mRNA cleavage depending partly on the level of complementarity between the small RNA and target (He et al, 2004). So, the gene silencing function of miRNA and siRNA is essentially related to a code existed in the 21-25 nucleotide RNA sequence. The example also shows the formation of any code relation is a very complex process which needs the participation of many factors. Splice junctions at the exon-intron boundaries have a code GU at the 5' end and AG at the 3' end of the intron. But, to decide that GU is really a 5'splice site and AG is really a 3' splice site in human genes a consensus sequence A/CAGGURAGU around GU and a consensus sequence YNYURAY near branching point A， followed by polyprimidine Y10-20 and then YAG around 3' splice site are required (Clark et al, 2002; Zhang et al, 2004). These consensus sequences around splice sites form the splice code. The five small nuclear RNA (snurps) – U1,U2,U5,U4 and U6, together with 50-100 additional proteins, form the splicesome responsible for the lariat-like splicing mechanism. The splice code indicates the location in pre-RNA sequence where the splicesome can bind and the splicing mechanism can work. Recently, it is known that 75%of human genes have alternative splicing. There are five alternative splice forms, namely intron retention, alternative donor introns, alternative acceptor introns, cassette exons and mutually exclusive exons. All these splice forms can be understood based on the interaction between splicesome and splice code sequence. The alternative splicing leads to a considerable increase of protein products, so it is one of the most significant components for the functional diversity in human genome evolution.

Generally, the extension of code is dependent on three-element-interaction among DNA,RNA and protein. However, evolution is a mender; sometimes the modes for coding have been beyond the scope of DNA, RNA and protein interaction. In eukaryotes the chromatin remoulding (i.e., the change of chromatin structure) is an important event for controlling gene expression. Histone octamer in nucleosome can be modified by methylation, acetylation and phosphorylation. They alter the local chromatin structure and activate the gene. Recently it was found in human genome the transcription starts are largely influenced by the feature of chromatin accessibility and histone modification (ENCODE, 2007). The chromatin remoulding and the wrapping DNA molecule form two co-linear sequences of information source. Interestingly, once some histone modifications established, such changes in chromatin may persist through cell division, creating an epigenetic state in which the properties of a gene are determined by the self-perpetuating structure of chromatin. The name epigenetic reflects the fact that a gene may have an inherited condition (it may be active or may be inactive) that does not depend on its DNA sequence. The self-perpetuating structure of prions is an example of epigenetic heredity. The common occurrence of epigenetic state in mammalian heredity reflects the evolutionary extension of the mode for coding.

Gene transfer from the outside is another approach to the growing of coding information quantity of a genome. It includes horizontal gene transfer (lateral gene transfer) and symbiotic mergers between species. The horizontal transfer means a gene is copied from the genome of one species into that of another species. It is probably frequent in bacteria. We know that about 24% *B. subtilis* genes have clear orthologous counterparts in *E. coli,* and 66% *H. influenzae* proteins have homologues to *E. coli* proteins. However, the detection of a horizontal-transfer event should be made based on rigorous phylogenetiic analysis or by either recombination or gene conversion (Li, 1997). Genes are even known to have transferred between Archaea and bacteria. Genes probably also occasionally transfer from bacteria into multicellular eukaryotes. For example, there are 223 proteins in human genome that have significant similarity to proteins from bacteria. At least 113 of these genes appear to be present only in vertebrates. But, whether they are horizontal-transfer event from bacteria to vertebrates remains to be tested.(IHGSC,2001). The symbiotic mergers means two species combine their genomes into one in a particularly intimate symbiosis. It is argued that 2000-2500 million years ago the symbiosis between two bacteria led to the eukaryotic cell containing a mitochondrion. The newly merged cell might have had two DNA molecules and one of them has expanded and evolved into the nuclear DNA and another shrunk and evolved into the mitochondrial DNA (Ridley, 2004).

## 4. Conclusions

Different from energy, information is not conservative during its transmission. The information expansion is a basic law in biology with its meaning like energy conservation in physics. A great deal of molecular biological data show the information quantity gradually expanded in evolution. DNA sequence duplication is the first important factor for the expansion. Sequence duplication includes short fragment repetition, regional multiplication, gene and genome duplication, etc.. New DNAs are produced and maintained in duplications just by their ability to replicate and these duplicated sequences greatly contribute to the genome information expansion. Enlargement of genetic information network is another important factor. Life is an information system of three-element-interaction among DNA, RNA and protein. The formation, storage, expression and transmission of genetic information are generally realized in the interaction network of DNA,RNA and protein. For eukaryotes the chromatin remoulding and histone modifications is another element entering in the information interaction network. A Chinese ancient philosopher Laozi said: “One generates two, two generates three and three generates all things in the Universe.” So, *three* in this sense means *infinity* (Three＝Infinity). Three- or more-element-interaction opens up more possibilities and shows more complexities than two-element-interaction. The interaction network of multi-elements makes the genetic language complex enough to be able to represent the life. For example, RNA and protein can have arithmetic function operated on DNA sequence, modifying, deleting or inserting some segments on it and making the function-coding information of genome increasing. The regulation of gene expression is realized through proteins reacting on DNA in usual regulatory mechanism; it can also be realized through ncRNA interaction in three-element-interaction network. In fact, as early as Jacob and Monod proposed the regulatory model they suggested the possibility of the highly specific interaction between operon RNA and transcripts of regulatory gene. The recent discovery of histone modifications that strongly influence the transcription starts in human genome provides another example of how the genetic information network is enlarged. Therefore, the duplication of DNA sequence and the enlargement of genetic information network centered at DNA sequence are two internal predominating factors in coding information quantity expansion of a genome.

However, the change of information quantity of a genetic system depends on environment and many stochastic factors. Both DNA expansion and contraction have been found in genome evolution. Only after taking the functional selection into account the directionality of DNA evolution can then be established. That is, although there exist a lot of accidental factors changing the genome size in the course of evolution, the global trend of the function extension of the genome that makes the species more suitable to the environment and more competitive to win is decisive and in single-direction. The central points of Darwin's evolutionary theory are adaptation and competition. Following Darwin, as environments change, and competing species change, species will evolve new adaptations. The competition between similar individuals will make for the evolution of new adaptations in each that reduce the intensity of competition; divergence will thus result. This explains why phylogeny is always hierarchical and tree-like (tree-like means directionality) and how the evolutionary directionality emerges. Here the key point is functional selection. A new species formation is at the cost of the failure of many trials of old species. The advantageous function acquired by a species makes it becoming a winner. So, the directionality of biological evolution is easily understood in this way. What we proposed in the article is: the coding information quantity of a genome that codes for the function of species can serve as the measure of directionality. The needs of function is the motive force for the expansion of coding information quantity and the information quantity expansion is the way to make functional innovation and extension. So, the increase of coding information quantity of a genome is a measure of the acquired new function and it determines the directionality of genome evolution. We know that there are several time arrows in physics which express the irreversibility of the physical laws. The proposed directionality here is a new time arrow but it should be understood only in terms of biology. As is well known, physics used to ask "why", while biology used to ask "what for". Purpose of function improvement is fully a biological concept. Based on the functional selection we are able to find a measure for evolutionary directions, namely the function-coding information quantity growing in evolution. Although the CIQG law proposed here still needs further quantitative demonstrations we have found the law consistent (or at least not in conflict) with all up-to-date experimental data in genomics.

The practical meaning of this study is: it will be helpful to understand the expansion of the coding information quantity in genome evolution and especially, to understand the distribution of 'junk' DNA as a kind of 'dark information' occurred in eukaryotic genomes and to find the possible new code relations existed therein. The life information emerges from the stochastic background of inanimate Nature through millions of years' natural selection, condensed from a huge amount of accidents. What is the main line of the present theoretical molecular biology? We proposed that the main line should be focused on the flow of life information, that is, on the fundamental laws in heredity, transmission, regulation and expression of genetic information. We called the main line as information biology (Luo, 2005;2006). It is expected that the suggested CIQG law will be a first cornerstone of information biology.

The paper is a revised version of arXiv: q-bio/0805.1085(2008) under the title "On the Law of Directionality of Genome Evolution".

**Acknowlegement** The work was supported by National Science Foundation of China, No 90403010

## References

Aury JM, Olivier J, Duret L et al. Global trends of whole-genome duplications revealed by the ciliate *Paramecium tetraurelia*. Nature 2006, **444**:171-178.

Bejerano G., Pheasant M., Makunin I., Stephen S., Kent WJ., Mattick JS., Haussler D. Ultraconserved elements in the human genome. Science, 2004, **304**:1321-1325.

Bennett MD., Leitch IJ. Genome size evolution in plants. In: *Evolution of the Genome* (Edi. Gregory TR), Elsevier Inc 2005.

Bininda-Edmonds, Cardillo M, Jones KE et al. The delayed rise of present-day mammals. Nature 2007, **446**:507-512.

Charlesworth B., Sniegowshi P., Stephan W. The evolution dynamics of repetitive DNA in eukaryotes. Nature 1994, **371**:215.

Chou HH, Hayakawa T, Diaz S et al. Inactivation of CMP-N- acetylneuraminic acid hydroxylase occurred prior to brain expansion during human evolution. Proc. Natl. Acad. Sci USA 2002, **99**:11736-11741.

Clark F, Thanaraj T A. Categorization and characterization of transcript confirmed constitutively and alternatively spliced introns and exons from human. Human Molecular Genetics, 2002, **11**(4):451-464.

Dermitzakis E.T., Reymond A., Scamuffa N., Ucla C., Kirkness E., Rossier C., Antonarakis S. E. Evolutionary discrimination of mammalian conserved non-genic sequences[CNGs]. Science, 2003, **302:**1033-1035.

Des Marais DL, Ransher MD. Escape from adaptive conflict after duplication in an anthocyanin pathway gene. Nature 2008, **454**:762-765.

Devos KM, Brown JKM, Bennetzen JL. Genome size reduction through illegitimate recombination counteracts genome expansion in Arabidopsis. Genome Research 2002, **12**:1075-1079.

Enard W., Khaitovitch P. Klose J. et al. Intra- and interspecific variation in primate gene expression patterns. 2002, Science **296**:340-343.

Filkowski J, Kowalchuk O, Kowalchuk I. Dissimilar mutation and recombination rates in Arabidopsis and tobacco. Plant Sci. 2004, **166:**265-272.

Gregory TR. Genome size evolution in animals. In: *Evolution of the Genome* (Edi. Gregory TR), Elsevier Inc 2005.

Gregory TR. DeSalle R. Comparative genomics in prokaryotes. In: Evolution of the Genome (Edi. Gregory TR), Elsevier Inc 2005.

Gregory TR et al. :Eukaryotic genome size databases. Nucleic Acids Research 2007, **35,** Database issue D332-D338.

International Human Genome Sequencing Consortium. Initial sequencing and analysis of the human genome. Nature 2001, **409**:860-921.

Kazazian HH. Mobile elements: drivers of genome evolution. Science, 2004, **303:**1626-1632.

Kouzarides T. Chromatin modifications and their function. Cell, 2007, **128:** 693-705.

He L., Hamm G.J. MicroRNAs: Small RNAs with a big role in gene regulation. Nature Rev. Genetics, 2004, **5**: 522.

Leitch IJ., Bennett MD. Genome downsizing in polyploidy plants. Biol. J. Linn. Soc. 2004, **82**:651-663.

Lewin, B. *Gene VIII*. Pearson Education Inc., 2004; *Gene IX*. Jones & Bartlet Publishers, Inc.,2008.

Li WH. *Molecular Evolution.* Massachusetts: Sinauer Associates 1997.

Liu G., Eichler E. et al. Analysis of primate genomic variation reveals a repeat-driven expansion of the human genome. Genome Res. 2003, **13**: 358-368.

Luo LF. *Physical Aspects on Life Evolution.* (in Chinese) Shanghai Science &Technology Pub 2000.

Luo LF. Information biology – an introduction. Journal of Inner Mongolia University 2005, **36**: 653-664.

Luo LF. Junk DNA and information biology. Science (in Chinese) 2006, **58**:24-28.

Makalowski W. Not junk after all. Science, 2003, **300**:1246-1247.

Mira A, Ochman H, Moran NA. Deletional bias and the evolution of bacterial genomes. Trends Genet. 2001, **17**:589-596.

Nusbaum C et al. DNA sequence and analysis of human chromosome 18. Nature, 2005, **437**:551-555.

Organ CL, Shedlock AM, Meade A, Pagel M, ,Edwards SV. Origin of avian genome size and structure in non-avian dinosaurs. Nature 2007, **446**:180-184.

Ozkan H, Levy AA, Feldman M. Allopolyploidy-induced rapid genome evolution in the wheat group. Plant Cell 2001, **13**: 1735-1747.

Peters SE. Environmental determinants of extinction selectivity in the fossil record. Nature 2008, **454**: 626-629.

Petrov DA,, Sangster TA., Johnston JS., Hartl DL., Shaw KL. Evidence for DNA loss as a determinant of genome size. Science 2000, **287**:1060-1062.

Patthy L. *Protein Evolution*. Oxford: Blackwell Science 1999.

Ridley M. *Evolution* (3rd edition). Blackwell Publishing 2004.

Scannell DR, Byrne KP, Gordon JL, Wong S, Wolfe KH. Multiple rounds of speciation associated with reciprocal gene loss in polyploidy yeasts. Nature 2006, **440**: 341-345.

Schrodinger E. *What is Life*? Cambridge : Cambridge Univ. Press.1944.

Storz G. An expanding universe of noncoding RNAs. Science, 2002, **296**:1260-1263.

Stadtman TC. Selenocyteine. Ann Rev. Biochem. 1996, **65** 83-100.

Smit AFA. Interspersed repeats and other momentos of transposable elements in mammalian genomics. Curr. Opin. Genet. Dev. 1999, **9**:657-663.

The ENCODE Project Consortium. Identification and analysis of functional elements in1% of the human genome by the ENCODE pilot project. Nature 2007, **447**:799-816.

Wapinski I, Pfeffer A, Friedman N, Regev A. Natural history and evolutionary principles of gene duplication in fungi. Nature 2007, **449**: 54-61.

Winter H, Langbein L, Krawczak M et al. Human type 1 hair keratin pseudogene. Human Genet. 2001, **108**:37-42.

Wickelgren I. Spinning junk into gold. Science, 2003, **300**:1646-1649.

Xu X, Norell MA. A new troodontid from China with avian-like sleeping posture. Nature 2004, **431**: 838-841.

Zhang LR., Luo LF. Splice site prediction with quadratic discriminant analysis using diversity measure. Nucleic Acids Research 2003, **31**:6214-6220.